\documentclass[aps,prl,12pt,a4paper,twocolumn,groupedaddress,preprintnumbers,floatfix,nofootinbib,showpacs]{revtex4}
\usepackage[english]{babel}
\usepackage{amsmath}
\usepackage{graphicx}
\usepackage{color}
\usepackage{amssymb}
\usepackage{hyperref}
\usepackage{dcolumn}
\usepackage{bm}
\usepackage{graphicx}
\usepackage{float}

\def\lnv{lepton number violation }
\def\vev#1{\left\langle #1\right\rangle}

\def\e6{$\mathrm{E(6)}$ }
\def\10{$\mathrm{SO(10)}$ }
\def\21{$\mathrm{SU(2)_L \otimes U(1)_Y}$ }
\def\31{$\mathrm{SU(3)_c \otimes U(1)_Q}$ }
\def\SM{$\mathrm{SU(3)_c \otimes SU(2)_L \otimes U(1)_Y}$ }
\newcommand{\sm}{{Standard Model }}

\def\3211{$\mathrm{SU(3) \otimes SU(2)_L \otimes U(1)_R \otimes U(1)_{B-L}}$ }
\def\321{$\mathrm{SU(3) \otimes SU(2) \otimes U(1)}$ }
\def\422{$\mathrm{SU(4) \otimes SU(2) \otimes SU(2)_R}$ }

\def \znbb {$0\nu\beta\beta$ }

\newcommand{\AddrAHEP}{%
  AHEP Group, Institut de F\'{i}sica Corpuscular --
  C.S.I.C./Universitat de Val\`{e}ncia, Parc Cientific de Paterna.\\
 C/ Catedratico Jose Beltran, 2 E-46980 Paterna (Valencia) - SPAIN}

\newcommand{\AddrWurzburg}{%
  Institut f{\"u}r Theoretische Physik und Astrophysik,\\
  Universit{\"a}t W{\"u}rzburg, 97074 W{\"u}rzburg, Germany}

\begin{document}
\preprint{DCP-13-03}

\title{Dirac neutrinos from flavor symmetry}

\author{
Alfredo Aranda,$^{1,2}$\footnote{Electronic address:fefo@ucol.mx} 
Cesar Bonilla,$^{3}$\footnote{Electronic address:cesar.bonilla@ific.uv.es}
S. Morisi,$^{4}$\footnote{Electronic address:stefano.morisi@gmail.com}
E. Peinado,$^{5}$\footnote{Electronic address:eduardo.peinado@lnf.infn.it}
J. W. F. Valle,$^{3}$\footnote{Electronic address:valle@ific.uv.es}}

\affiliation{
$^1$Facultad de Ciencias - CUICBAS,
Universidad de Colima, Colima, Mexico.\\
$^2$Dual C-P Institute of High Energy Physics, Mexico.\\
$^3$ \AddrAHEP \\
$^4$ \AddrWurzburg \\
$^5$ INFN, Laboratori Nazionali di Frascati, Via Enrico Fermi 40, I-00044 Frascati, Italy.}
\date{\today}

\begin{abstract} 
  We present a model where Majorana neutrino mass terms are forbidden
  by the flavor symmetry group $\Delta(27)$.  Neutrinos are Dirac
  fermions and their masses arise in the same way as those of the
  charged fermions, due to very small Yukawa couplings. The model fits
  current neutrino oscillation data and correlates the octant of the
  atmospheric angle $\theta_{23}$ with the magnitude of the lightest
  neutrino mass, with maximal mixing excluded for any neutrino mass
  hierarchy.
\end{abstract}
\maketitle

\noindent
\textbf{Introduction}

\vskip2.mm

The historic observation of neutrino
oscillations~\cite{art:2012,An:2012eh, Ahn:2012nd, Abe:2011sj} implies
that neutrinos are massive in contrast with the Standard Model (SM)
prediction. Incorporating small masses requires an extension of the SM
in which neutrinos are generally expected to be of Majorana type,
hence violating lepton number
symmetry~\cite{Schechter:1980gr}~\footnote{Recently it has been
claimed that one can find models where lepton number is violated by
four units, $\Delta L=4$, even if neutrinos are of the Dirac
type~\cite{Heeck:2013rpa}.}. On the other hand in many schemes, such
as for example the so-called seesaw mechanism \lnv is expected to
account for the observed smallness of neutrino mass relative to that
of charged fermions~\cite{Schechter:1980gr}. Yet, so far current
neutrino oscillation experiments have been insensitive to the Majorana
nature of neutrinos~\cite{Schechter:1982bd,Duerr:2011zd} and, despite
intense ongoing efforts it has not been confirmed through the
observation of \lnv processes such as neutrinoless double beta decay
(\znbb)~\cite{Barabash:2011fn}. Hence neutrinos could very well be
Dirac fermions~\cite{Memenga:2013vc}. In short, the status of lepton
and baryon number symmetries remains as one of the deepest unsolved
mysteries of nature~\cite{weinberg:1980bf}.
An equally puzzling challenge is associated to the origin of the
peculiar flavor pattern of mixing angles indicated by global fits of
neutrino oscillation experiments~\cite{Tortola:2012te}.

Here we suggest a possible interconnection between these puzzles,
namely, that lepton number conservation can be an accidental
consequence of the flavor symmetry that accounts for the neutrino
mixing pattern.

Over the last decade non-Abelian discrete groups have been widely used
as family symmetries because of their potential in restricting
neutrino mixing patterns~\cite{Morisi:2012fg,Hirsch:2012ym}. As
examples we mention the successful models based on
the $A_{4}$ group predicting $\theta_{23}=\pi/4$ and
$\theta_{13}=0$~\cite{babu:2002dz,altarelli:2005yp}.
However the recent discovery of a large reactor angle,
$\theta_{13}>0$~\cite{An:2012eh, Ahn:2012nd, Abe:2011sj}, and a
possible hint in favor of non-maximal atmospheric mixing present in
recent oscillation fits $\theta_{23}$~\cite{Tortola:2012te} suggests
the need for generalizing these models~\cite{Morisi:2013qna} and/or
seeking for alternative schemes based upon different flavor
symmetries~\cite{Ding:2012wh}.

Here we present a flavor model for leptons using the non-Abelian group
$\Delta(27)$~\cite{Ma:2006ip,deMedeirosVarzielas:2006fc,Ma:2007wu,Morisi:2012hu}
that is able to provide automatic lepton number conservation in a way
consistent with current global fits of neutrino oscillation
data~\cite{Tortola:2012te}. Recently other non-Abelian flavor
symmetries have been used for pure Dirac neutrinos, see for
instance~\cite{Chen:2012jg, Ding:2013eca, Holthausen:2013vba,
  Memenga:2013vc}, however Majorana mass terms are forbidden by means
of extra Abelian symmetries. Here we focus on the possibility that
Majorana mass terms are not allowed from the flavor symmetry without
requiring any extra additional Abelian symmetry.  We note that since
neutrinos are Dirac fermions \znbb is exactly forbidden.
In addition the model gives a correlation between 
neutrino oscillation parameters that excludes the maximal
$\theta_{23}$ value. \\ 

\vskip3.mm

\noindent
\textbf{Preliminaries}

\vskip2.mm
In order to generate Dirac neutrino mass terms we introduce singlet
``right handed'' (RH) neutrinos transforming under the flavor symmetry
group $\mathcal{G}_{F}$ in such a way that their tensor product does
not contain the trivial element of $\mathcal{G}_{F}$.
This means that, even though lepton number conservation is not
necessarily required~\textit{a priori}, Majorana mass terms are
forbidden in the Lagrangian as a result of the flavor symmetry
$\mathcal{G}_{F}$.

Although this may be achieved by using an Abelian symmetry
$\mathbb{Z}_{N}$ $\forall$ $N\geq3$ our focus relies on simple non-Abelian
flavor symmetry groups.  We assume that RH-neutrinos ($N_R$)
transform as a 3-dimensional irreducible representation (irrep) of
$\mathcal{G}_{F}$.  Hence if $N_{R}$ transforms as $3$-dimensional
irrep ($\bf{3}$) under $\mathcal{G}_{F}$, one finds that the
non-Abelian symmetries which forbid a term like $N_{R}^T N_{R}$
are~

\begin{itemize}
\item $\Delta(3N^{2})$ for $N\geq3$: these groups contain nine
  singlets and $(N^{2}-3)/3$ triplets for $N=3\mathbb{Z}$.  Otherwise,
  for $N\neq3\mathbb{Z}$, they have three singlets and $(N^{2}-1)/3$
  triplets.
 \item$\Sigma(3N^{3})$ for $N\geq3$: the set of groups with
   $N(N^{2}+8)/3$ conjugacy classes, $3N$ singlets and $N(N^{2}-1)/3$
   triplets.
 \item $T_{N}$ for the $N$ values given in \cite{Ishimori:2010au}
   these groups have 3 singlets and $(N-1)/3$ three-dimensional
   irreducible representations.
 \item $Z_9\rtimes Z_3$.

\end{itemize}
In fact the mass term $N_{R}^{T}N_{R}$ is forbidden because the tensor product
  ${\bf 3}_{i}\otimes{\bf3}_{i}$ (where $i=1,...,n_{d}$ and 
  $n_{d}=(N^{2}-3)/3$ for $\Delta(3N^{2})$ and $n_{d}=N(N^{2}-1)/3$
  for $\Sigma(3N^{3})$) does not contain a trivial 1-dimensional
  irrep $1^{0}$~\cite{Luhn:2007uq,Ishimori:2010au}.

\vskip3.mm

\noindent
\textbf{The model}

\vskip2.mm

Searching for the the smallest realistic flavor symmetry group of the
above class, i.e. used in the context of forbidding Majorana mass
terms, we find that \footnote{$T_7$ has the desired product and has
  indeed been used as a successful flavor symmetry, however not in the
  context of Dirac
  neutrinos~\cite{Hagedorn:2008bc,Cao:2010mp,Parattu:2010cy,Luhn:2012bc}.}
it is $\Delta(27)$. The \SM$\otimes \ \Delta(27)$ multiplet assignment
is given in Table~\ref{tmc},\footnote{ We denote, by convenience,
  ${\bf1}\equiv{\bf1}_{(0,0)}$, ${\bf1}'\equiv{\bf1}_{(1,0)}$,
  ${\bf1}''\equiv{\bf1}_{(2,0)}$, ${\bf3}\equiv{\bf3}_{(0,1)}$ and
  ${\bf3}'\equiv{\bf3}_{(0,2)}$, where the index notation is that used
  in \cite{Luhn:2007uq,Ishimori:2010au}}.  where we have extended the
SM by adding three right-handed neutrinos and two Higgs doublets apart
from that of the \sm.
\begin{table}
\begin{tabular}{|c|c|c|c|c|c|c|}
\hline
      & $\overline{L}$& $\ell_{1R}$ & $\ell_{2R}$  & $\ell_{3R}$ & $N_{R}$ & $H$\\
\hline
 $SU(2)_{L}$& ${\bf2}$ & ${\bf1}$    &   ${\bf1}$  &  ${\bf1}$ &${\bf1}$ & ${\bf2}$\\      
\hline
 $\Delta(27)$& ${\bf3}$ & ${\bf1}$    &   ${\bf1}'$  &  ${\bf1}''$ &${\bf3}$ & ${\bf3'}$\\
 \hline
\end{tabular}\caption{Matter assignments of the model.}
\label{tmc}
\end{table}
The most general invariant Lagrangian for leptons is written as
\begin{equation}\label{Lyl}
\mathcal{L}_{\ell}=\sum_{i=1}^{3}Y_{i}^{\ell}\bar{L}\ell_{iR}H+Y^{\nu}\bar{L}N_{R}\tilde{H}+h.c.,
\end{equation}
where we use the compact notation $H = (H_1, H_2, H_3)$ and $\tilde{H} = (\tilde{H_1},\tilde{H_2},\tilde{H_3})$ with $\tilde{H_i}\equiv i\sigma_2H^*$. After electroweak symmetry breaking one gets the following
patterns for the neutrino and charged lepton mass matrices:
\begin{eqnarray}
    \label{eq:MlMnu}
M_{\nu}&=&
 \left[ \begin{array}{ccc}
a v_{1}  &      b v_{3}    &   c   v_{2} \\
c v_{3}  &     a v_{2}     &   b v_{1}\\
b v_{2}  &     c  v_{1}    &    a v_{3}
\end{array} \right] \\ \nonumber
M_{\ell}&=&
 \left[ \begin{array}{ccc}
Y_{1}^{\ell}v_{1}  &                Y_{2}^{\ell}v_{1}    &             Y_{3}^{\ell}v_{1} \\
Y_{1}^{\ell}v_{2}  & \omega         Y_{2}^{\ell}v_{2}     &  \omega^{2} Y_{3}^{\ell}v_{2}\\
Y_{1}^{\ell}v_{3}  & \omega^{2}     Y_{2}^{\ell}v_{3}    &  \omega     Y_{3}^{\ell}v_{3}
\end{array} \right]
\end{eqnarray}
where $v_{i}$ are Higgs scalar vacuum expectation values (vevs),
$\vev{ H}=(\vev{ H_{1}}, \vev{
  H_{2}},\vev{H_{3}})=(v_{1},v_{2},v_{3})$. The parameters
$\{a,b,c,Y_{i}\}$ are real if CP invariance is assumed where the CP
transformation is properly defined in
\cite{Branco:1983tn,Ferreira:2012ri, Holthausen:2012dk,Nishi:2013jqa}.
One sees that in such minimal scenario the smallness of neutrino
masses w.r.t. those of the charged leptons must arise due to very
small Yukawa couplings \footnote{ Suppressed Yukawa coefficients can
  arise in extra dimension schemes, i.e. \cite{Ding:2013eca}, as well
  as supersymmetric schemes, see for instance \cite{Chen:2012jg}}.
The structure of $M_{\ell}$ and $M_{\nu}$ are well known in the 
literature~\cite{Ma:2006ip,altarelli:2005yp} and the alignment $\vev{H}=v(1,1,1)$ turns out to be natural in $\Delta(27)$~\cite{Ma:2006ip,Ma:2007wu}. 

In such a case $M_{\ell}$ can be written as $M_{\ell}=U_{\omega}\hat{Y}$ where
$\hat{Y}=\text{diag}(Y_{1},Y_{2},Y_{3})$ and
\begin{equation}
U_{\omega}=\frac{1}{\sqrt{3}}\left[ \begin{array}{ccc}
1  &       1            &     1 \\
1  & \omega            &  \omega^{2} \\
1  & \omega^{2}        &  \omega     
\end{array} \right]
\end{equation}
is the so--called ``magic'' matrix.  However, given the structure of
the neutrino mass matrix $M_{\nu}$, the previous alignment
$\vev{H}=v(1,1,1)$ cannot be assumed since then $U_{\omega}$
diagonalizes both $M_{\nu}M_{\nu}^\dagger$ and
$M_{\ell}{M_{\ell}}^\dagger$. This results in a trivial lepton mixing
matrix
\begin{equation}
U=U_{\ell}^{\dagger}U_{\nu}= U_{\omega}^{\dagger}U_{\omega}=\mathbb{I}.
\end{equation}
Moreover, when $v_{1}=v_{2}=v_{3}=v$ and the couplings $a$, $b$ and
$c$ are real the resulting neutrino masses are also not suitable to
account for current neutrino oscillation data.

All of this can be avoided by deviating from the simplest vev
alignment, i.e. we can fit the neutrino squared mass differences, as
well as induce large lepton mixing angles by assuming that the vev
alignment is generalized to
\begin{equation}\label{al2eps}
\langle H \rangle=\hat{v}(1+\epsilon_{1},1+\epsilon_{2},1)^{T} \ ,
\end{equation}
where $|\langle H \rangle|^2 = v^2 = (246 \ \rm{GeV})^2$. The above vev configuration is a solution of the minimization
conditions of the scalar potential provided it softly breaks the
flavor symmetry, the deviation parameters $\epsilon_{1,2}$ being then
associated to this soft breaking.

Taking into account Eq.~(\ref{al2eps}) the mass matrices for the
lepton sector are now given by
\begin{eqnarray}\label{Mseps}
M_{\nu}&=&\hat{v}\left[ \begin{matrix}
a (1+\epsilon_{1})  &      b     &   c   (1+\epsilon_{2}) \\
c   &     a (1+\epsilon_{2})     &   b (1+\epsilon_{1})\\
b (1+\epsilon_{2})  &     c  (1+\epsilon_{1})    &    a 
\end{matrix} \right]  \\ \nonumber
M_{\ell}&=&\hat{v}
 \left[ \begin{matrix}
Y_{1}^{\ell}(1+\epsilon_{1})  &                Y_{2}^{\ell}(1+\epsilon_{1})    &             Y_{3}^{\ell}(1+\epsilon_{1}) \\
Y_{1}^{\ell}(1+\epsilon_{2})  & \omega         Y_{2}^{\ell}(1+\epsilon_{2})     &  \omega^{2} Y_{3}^{\ell}(1+\epsilon_{2})\\
Y_{1}^{\ell}  & \omega^{2}     Y_{2}^{\ell}    &  \omega     Y_{3}^{\ell}
\end{matrix} \right]\nonumber. 
\end{eqnarray}
Note that an immediate consequence of the generalized vev alignment is
that the $U_{\omega}$ no longer diagonalizes the neutrino mass matrix
nor that of the charged leptons, and therefore, as desired, the lepton
mixing matrix is now non-trivial,
\begin{equation}
U=U_{\ell}^{\dagger}U_{\nu}\ne \mathbb{I} \ .
\end{equation}
Furthermore one can indeed fit all neutrino observables as we now
show.

\vskip3.mm
\noindent
\textbf{Results}

\vskip2.mm

\begin{figure}[h]
               \centering
                \includegraphics[width=8cm]{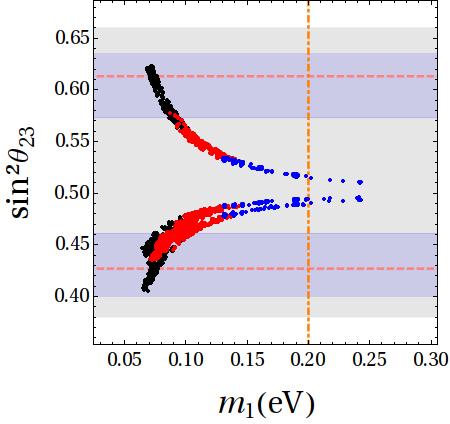}
                \caption{Correlation between the atmospheric angle and
                  the lightest neutrino mass for the NH case. The
                  horizontal dotted lines represent the best fit
                  values, the (dark) blue and (light) gray horizontal
                  bands are the $1\sigma$ and $2\sigma$ allowed
                  ranges, respectively. The blue (light gray), red
                  (gray) and black points are model expectations
                  corresponding to vev deviations of $10\%$, $20\%$
                  and $30\%$ respectively (see text for more
                  details). The vertical dot-dashed line indicates
                  KATRIN's sensitivity~\cite{Bornschein:2003xi}.  }
         \label{23-m1-nh}
         \end{figure}
Here we consider deviations of the alignment $v(1,1,1)$ of the order
$\mathcal{O}(\lambda_C)$ where $\lambda_C\sim 0.2$ is the Cabibbo
angle. More precisely, using Eqs.~(\ref{al2eps}) and (\ref{Mseps}) we
have scanned over values for the small parameters $\epsilon_{1,2}$
within the range $|\epsilon_{1,2}|\leq0.3$\, and selected those
solutions which satisfy the global fits for the mixing angles at
$3\sigma$~\cite{Tortola:2012te}
$$ 
 0.017 <\sin^2\theta_{13}<0.033 $$
$$0.36(0.37) <\sin^2\theta_{23}<0.68(0.67)~\text{NH(IH)}$$
$$0.27 <\sin^2\theta_{12}<0.37,$$
as well as the neutrino squared mass differences 
\begin{small}
\begin{eqnarray}
\Delta m_{21}^2&=&(7.12-8.20) \times 10^{-5} eV^{2},\nonumber\\
|\Delta m_{31}^2|&=&
\left\lbrace\begin{array}{cc}
         (2.31-2.74) & \text{for NH}\nonumber \\
        (0.21-2.64) & \text{for IH} \nonumber
\end{array}
\right\rbrace \times 10^{-3} eV^{2} \ .
\end{eqnarray}
\end{small}

\begin{figure}[h]
               \centering
                \includegraphics[width=8cm]{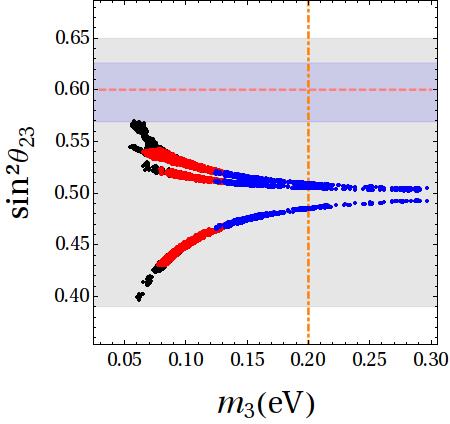}
                \caption{Same as Figure~\ref{23-m1-nh} for the IH
                  case. Note that in this case a $30\%$ vev deviation
                  is not enough to reach the best fit value of
                  $\theta_{23}$.}
         \label{23-m1-ih}
         \end{figure}

We have found a correlation between the atmospheric angle and the
lightest neutrino mass for both the normal mass hierarchy (NH) and the
inverted mass hierarchy (IH) cases. This is shown in
Figures~\ref{23-m1-nh} and \ref{23-m1-ih} for the NH and IH cases,
respectively. In both figures the dotted horizontal lines represent
the best fit values, while the (dark) blue and (light) gray horizontal
bands are the $1\sigma$ and $2\sigma$ bands obtained in
Ref.~\cite{Tortola:2012te}, respectively. For the NH case the global
oscillation fit finds also a local minimum in the first octant of
$\theta_{23}$~\cite{Tortola:2012te}.

         In order to explore the sensitivity of the observables with
         respect to the values of the vev deviation parameters,
         $\epsilon_{1,2}$, we consider the following cases,
         $|\epsilon_{1,2}|\lesssim 0.1$, $|\epsilon_{1,2}|\lesssim
         0.2$ and $|\epsilon_{1,2}|\lesssim 0.3$ where each one
         represents deviations of $10\%$, $20\%$ and $30\%$
         respectively.  As we mentioned above, the free parameters
         $\epsilon_{1,2}$ are associated to the $\Delta(27)$-soft
         breaking terms in the scalar potential and they are allowed
         to deviate at most at the order of the Cabibbo angle,
         $\epsilon_{1,2}\sim \mathcal{O}(\lambda_C)$.

         The solutions in blue (light gray) correspond to deviations
         up to $10\%$, those in red (gray) up to $20\%$ and those in
         black up to $30\%$. Figure~\ref{23-m1-ih} for the IH case
         shows that a $30\%$ vev deviation is not enough to reach the
         best fit value for $\theta_{23}$, so that larger deviations
         would be required in order to accomplish it.

         In the near future the KATRIN experiment could discover a
         neutrino mass in the degenerate region, going from
         $m_{\beta}\sim 0.3$~eV at $3\sigma$ significance to
         $m_{\beta}= 0.35$~eV at $5\sigma$
         significance~\cite{Bornschein:2003xi}.  If a neutrino mass is
         not seen in tritium $\beta$ decays this will set an upper
         bound of $0.2$~eV for neutrino mass and such a bound is
         depicted in each figure with the dot-dashed vertical line.

         It is important to note that the atmospheric angle deviates
         significantly from the maximal value as the vev deviations
         increase.

         Before concluding we mention that the model leads to
         contributions to flavor changing neutral current (FCNC)
         processes in the lepton sector, such as $\mu \to e
         \gamma$. However, we have checked a few representative points
         with normal neutrino mass hierarchy, and found that there is
         sufficient freedom in parameter space to satisfy the current
         MEG bound for such a process~\cite{Adam:2013mnn}. Indeed,
         Table~\ref{tbrs} gives the expected $\mu \to e \gamma$
         branching ratios such points are all consistent with current
         bounds. Considering that these points are located in different
         parameter regions, we believe that a detailed analysis will
         give similar results, though a complete study is beyond the
         scope of this paper and will be considered elsewere.
       Note that the model does not lead to FCNC in the quark sector
       as its symmetry affects only the lepton sector. A model
       upgrading the flavour symmetry to both sectors is being
       developed and will be presented in a future publication,
       including a detailed phenomenological study.
         \begin{table}[H]
         \centering
         \begin{small}
         \begin{tabular}{|c|c|c|c|}
         \hline
          Cases & $\text{Br}^{\text{th}}(\mu\to e\gamma)$ & $m_{\nu_{1}}$ (eV) & $\sin^{2}\theta_{23}$ \\ \hline
          i)   & $1.98\times 10^{-14}$ & 0.2399 & 0.4956  \\ \hline
          ii)  & $1.74\times 10^{-14}$ & 0.0930 & 0.4615  \\ \hline
          iii) & $1.65\times 10^{-14}$ & 0.0762 & 0.6107  \\ \hline
        \end{tabular}\caption{Theoretical branching ratios for the process $\mu\to e \gamma$ for three different cases corresponding to three different sets of ($\epsilon_{1}$,$\epsilon_{2}$), $m_{\nu_1}$, and $\sin^2\theta_{23}$. }\label{tbrs}
\end{small}
\end{table}
\vskip3.mm
\noindent
\textbf{Summary}

\vskip2.mm

We have presented a model based on $\Delta(27)$ flavor symmetry. We
showed that having RH neutrinos and LH leptons transforming as 3
dimensional irreps under $\Delta(27)$ forbids Majorana mass terms so
that neutrinos are {\it naturally} Dirac-type, just as all other \sm
fermions~\cite{Memenga:2013vc}. There is accidental lepton number
conservation in the model caused by gauge symmetry, as in the SM, and
it is present before and after EWSB. Furthermore, due to the particle
content of the model, we find that all higher order Weinberg-type
operators $LHLH(H^\dagger H)^n$ for $n=0,1,2,...$ that might yield a
Majorana mass are not allowed by the symmetry $\Delta(27)$ and there
are neither scalar singlets nor triplets to realize any diagram
(operator) in \cite{Ma:1998dn, Bonnet:2009ej, Bonnet:2012kz}.  This
scenario is able to fit the current data in the lepton sector and
establishes a correlation between the octant of the atmospheric angle
$\theta_{23}$ and the magnitude of the lightest neutrino mass
eigenvalue which may
be probed by coming experiments.\\[4mm]


\noindent
\textbf{Acknowledgments}
\vskip2.mm

We would like to thank Luis Lavoura and Christoph Luhn for comments on
the manuscript.  This work was supported by MINECO grants FPA2011-22975,
MULTIDARK Consolider CSD2009-00064, by Prometeo/2009/091 (Generalitat
Valenciana), by EU ITN UNILHC PITN-GA-2009-237920. A.A thanks support
from CONACYT and PROMEP. S.M. thanks DFG grant WI 2639/4-1. C.B. thanks
support from EPLANET.




\end{document}